\documentclass[12pt]{iopart}

\usepackage{graphicx}
\usepackage{epsfig}

\begin{document}

\linespread{1} \tolerance=10000 \hbadness=10000 \vbadness=10000

\title[Loss of Superfluidity in Bose-Einstein Condensate in an Optical Lattice]{Loss of Superfluidity in Bose-Einstein Condensate in an Optical Lattice with Cubic and Quintic nonlinearity}

\author{Priyam Das$^{1}$, Manan Vyas$^{2}$ and Prasanta K. Panigrahi$^{1,2}$}

 \address{$^{1}$ Indian Institute of Science Education and Research (IISER), Salt Lake, Kolkata - 700106, India}
 \address{$^{2}$ Physical Research Laboratory, Navrangpura, Ahmedabad - 380009, India}
           
\ead{daspriyam3@gmail.com}
 


\begin{abstract} 
In a one dimensional shallow optical lattice, in the presence of both
cubic and quintic nonlinearity, a superfluid density wave,
is identified in Bose-Einstein condensate. Interestingly, it ceases to
exist, when only one of these interaction is operative. We predict the
loss of superfluidity through a classical dynamical phase transition,
where modulational instability leads to the loss of phase
coherence. In certain parameter domain, the competition between
lattice potential and the interactions, is shown to give rise to a
stripe phase, where atoms are confined in finite domains.  In pure
two-body case, apart from the known superfluid and insulating phases,
a density wave insulating phase is found to exist, possessing two
frequency modulations commensurate with the lattice potential.  
\end{abstract}


\maketitle

\section{Introduction} The mean field Gross-Pitaevskii equation (GP)
\cite{Ginzburg,Pitaevskii} is known to capture the ground state
properties of the Bose-Einstein condensate (BEC) remarkably well, both
in three and lower dimensional configurations
\cite{Pethick,Dalfovo}. In the presence of an optical lattice
\cite{Morsch,Bloch1,Kui,Lin}, this picture changes drastically, when
quantum fluctuations become significant \cite{Subir}. For deep lattice
wells, fluctuation driven metal-insulator transition has been
experimentally confirmed \cite{Greiner,Jaksch,Liu,Das}. However, in
shallow wells, the quantum fluctuations are expected to be small,
making GP equation useful for identifying ground state structures.

For discrete nonlinear Schr\"odinger equation, Smerzi {\it et al.} in
2002 \cite{Smerzi}, predicted a dynamical superfluid insulator
transition (DSIT), which is different from the fluctuation driven
quantum phase transition.  Modulational instability, occurring in the
system, leads to the loss of phase coherence.  Hence, DSIT takes place
and the superfluid phase transits to the insulating state.  Followed
by this prediction, Cataliotti {\it et al.} in $2003$
\cite{Cataliotti}, reported the experimental observation of the
destruction of interference pattern and loss of superfluidity through
DSIT. In the same year, Adhikari demonstrated numerically the loss of
phase coherence and superfluidity of BEC \cite{Adhikari} with two-body
interaction.

In this paper, we demonstrate the existence of a superfluid density
wave state of BEC in an optical lattice, in the presence of both
cubic and quintic nonlinear interactions. This phase, with
twice the periodicity of the lattice, ceases to exist, when only
cubic and quintic nonlinearities are present. We have
considered two-body interaction to be repulsive and quintic
  interaction to be attractive for physical realization. The
existence of stripe phase is shown, where superfluid matter is found
only in finite domains. We predict that DSIT occurs in this system
also, where all the atoms transit from the superfluid phase to an
insulating phase, with analogous periodicity.  The filling fraction
in this case, varies inversely with the lattice site $n$ and
reaches an asymptotic constant value. It differs for odd and even
lattice positions with respect to a given site, taking a constant
value for even $n$. Interestingly, for odd $n$, it depends on the
depth of the lattice potential, as well as the interaction
strengths. In the presence of quintic nonlinearity alone, a
sinusoidal excitation is found, which exists only in the superfluid
phase, of similar periodicity as that of the optical lattice. In all
these cases, the dispersion plays a sub-dominant role. In presence of
pure two-body interaction, the superfluid density shows a periodicity
commensurate with the lattice. However, a density wave insulating
phase is found to exist, possessing two frequency modulations
commensurate with the lattice potential. The dynamical phase
transition connects the superfluid phase with an insulating phase
\cite{Carr}, which in turn is connected with the density wave
insulating phase.  The solutions are found to be marginally stable, as
per the Vakhitov-Kolokolov (VK) criterion \cite{Vakhitov}.


\section{BEC with cubic and quintic nonlinearity}
 The quintic nonlinearity in one dimension, corresponding to an
 effective three-body interaction, has appeared in a number of
 works. Muryshev {\it et al.}  \cite{Muryshev}, have investigated the
 dynamical stability and the dissipative dynamics of solitons in BEC,
 where the quintic nonlinearity arises in one dimension due to the
 interaction between axial and radial degrees of freedom. In
 \cite{Olshanii}, the cubic GP equation is used in one dimension, when
 the coupling constant is small, whereas, quintic interaction appears
 in the Tonks-Girardeau regime. The localization of ground state of
 BEC in optical lattice and the stability of BEC with quintic
 nonlinearity, have been analyzed by Abdullaev {\it et al.}
 \cite{Selerno,Gammal2}. Kolomeisky {\it et al.}, took a different
 approach and derived the GP functional with cubic and quintic
 nonlinearity for lower dimensions, using renormalization group method
 \cite{Kolomeisky1,Kolomeisky2}.

The quasi-1D GP equation in an optical lattice, in presence of both
cubic and quintic nonlinearity, takes the form
\cite{Muryshev,Olshanii,Gammal2,Selerno,Khaykovich},
\begin{equation}
i \psi_{t} = -\frac{1}{2}\psi_{zz} + ({g_{1}|\psi|^{2} + g_{2}}|\psi|^{4} + V(z) - \mu) \psi, \label{GPE}
\end{equation}
where, $V(z) = V_{0}\cos^{2}(z)$ is the optical lattice
potential. $g_{1}$ and $g_{2}$ are the appropriately normalized
strength of the cubic and quintic nonlinearity, respectively
\cite{Choi}. Since, both cubic and quintic terms are functions of
scattering length \cite{Khaykovich}, we describe the effect of the
interactions in terms of the ratio between cubic and quintic
nonlinearity $\kappa = g_{1}/g_{2}$.  The lattice and the chemical
potentials are normalized in terms of the recoil energy $E_{r}$. The
spatial co-ordinate and the wavefunction are scaled in units of
wavelength of incident laser light and $\sqrt{k}$ respectively; here
$k$ is the wave vector.

The ansatz solution of Eq. (\ref{GPE}), $\psi(z,t) = \sqrt{\sigma(z)} \exp{(i \chi(z) - i \omega t)}$ leads to a superfluid phase (SF):
\begin{eqnarray}
\psi_{SF} = \sqrt{\left(- \frac{g_{1}}{2 g_{2}} \pm
\sqrt{-\frac{V_{0}}{g_{2}}} \cos z\right)} e^{(i \chi(z) - i
\omega t)},\label{sol.2-3}
\end{eqnarray}
where, $\chi(z) = c \tan^{-1}[\sqrt{\frac{-g_{1}/2 g_{2} - \sqrt{-V_{0}/g_{2}}}{-g_{1}/2 g_{2} + 
\sqrt{-V_{0}/g_{2}}}} \tan(\frac{z}{2})]$, $\omega = \frac{1}{8} (1 - \frac{2 g_{1}^{2}}{g_{2}} - 
8 \mu)$ and $c = \pm \frac{1}{4 g_{2}} \sqrt{g_{1}^{2} + 4 g_{2} V_{0}}$. 
The periodicity of the density is twice that of the lattice potential, which indicates matter redistribution.
Atoms from neighbouring sites have been depleted leading to a density wave behaviour. As one can see from the 
expression, the phase vanishes where the superfluid density wave attains its maximum value. Since, the superfluid 
is characterized by a definite phase, the number of atoms on each lattice site is unknown, which allows the atom to move 
freely and it can easily tunnel from one lattice site to another.

Denoting $\frac{g_{1}}{2 g_{2}} \sqrt{-\frac{g_{2}}{V_{0}}}$ by $p$,
the solution exists only in the following regimes, when the quintic
nonlinear interaction and the lattice potential have opposite
signature:
\begin{enumerate}
\item{} If $\cos z > p$, then
\begin{description}
\item [a)] the solution exists for all $z$ for $p < - 1$, which
  indicates that the cubic and quintic nonlinearities are
  of opposite signature.
\item [b)] the solution does not exist for $p > 1$.
\item [c)] in the domain where, $-1 < p < 1$, the solution exists for $-\pi/2 < z < z_{j} < \pi/2$, where $z_{j} = \cos^{-1}p$.
This domain corresponds to a stripe phase, where superfluid matter is found only in finite domains.
\end{description}
\item{} If $\cos z < - p$, then
\begin{description}
\item [a)] the solution does not exist for $p > 1$.
\item [b)] the solution exists for all $z$ for $p < - 1$. Therefore,
  the cubic and quintic nonlinearities are of opposite
  signature.
\item [c)] when $-1 < p < 1$, the solution exists for $- \pi/2 < z < - z_{j} < \pi/2$, where $z_{j} = \cos^{-1}(p)$.
This also corresponds to a stripe phase.
\end{description}
\end{enumerate}

The average number of atoms per lattice site, the filling fraction, is found to be
\begin{equation}
\nu_{SF} = \frac{1}{n \pi}\int^{(n + 1/2)\pi}_{\pi/2} |\psi|^{2} dz = \sqrt{- \frac{V_{0}}{g_{2}}}
[ - p \pm \frac{1}{n \pi}(\cos(n \pi) - 1)].
\end{equation}

\begin{figure}[h]
\begin{center}
\includegraphics[scale = 0.5]{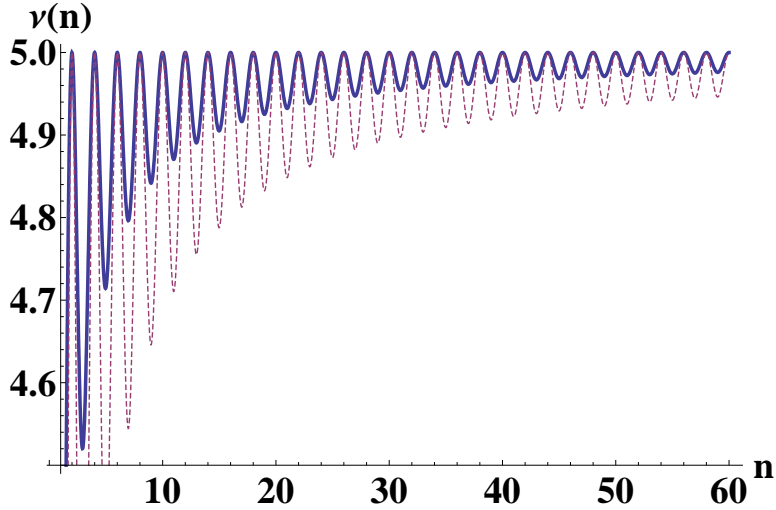}
\caption{The behaviour of the filling fraction when both cubic and
  quintic nonlinear interactions are operative.  $n$ is the lattice
  location with respect to a given site. The solid curve shows the
  filling fraction for the superfluid phase and the dotted curve is
  for the insulating phase with $V_{0} = 0.002$ (unit of recoil
  energy) and $\kappa = 5$.}
\label{fig2}
\end{center}
\end{figure}

Starting from a given lattice point, $\nu_{SF}$ depends on the lattice
location $n$, as depicted in Fig.(\ref{fig2}).  If $n$ is odd, the
filling fraction is a function of the lattice potential and varies
inversely with the lattice site number $n$. If $n$ is even, the
filling fraction takes a constant value depending only on the
interactions. The different behaviour of the filling fraction at the
neighbouring sites can be traced to the presence of $\cos (z)$ term in
the density. The odd-even effect arises due to the interplay between
the cubic and quintic nonlinear interactions, where a density wave is
present. As will be seen below, for the pure cubic or quintic
nonlinearity, this type of behaviour is absent.

The energy of the superfluid component is given by,
\begin{eqnarray}
E_{SF} &=& \frac{\pi g_{1}}{6 g_{2}^{2}}\left[g_{1}^{2} - 3 g_{2}
(2 + V_{0} - 2 \mu)\right] \nonumber \\ & & +  2 \pi\left(\frac{1}{2
g_{2}}\sqrt{g_{1}^{2} + 4 g_{2} V_{0}} - \frac{2 g_{2}
c^{2}}{\sqrt{g_{1}^{2} + 4 g_{2} V_{0}}}\right), \label{energy23}
\end{eqnarray}
which possesses a branch cut at, $V_{0} = - \frac{g_{1}^{2}}{4 g_{2}}$,
where the supercurrent $(J = c)$ vanishes. At this value, the phase coherence is lost and the superfluid phase
transits into an insulator (I) phase, which is driven by a modulational instability \cite{Smerzi}. 
The insulating wavefunctions with identical periodicity as that of the SF phase, have the following form:
\begin{eqnarray}
\psi_{I_{1}}(z,t) &=& \sqrt{- \frac{g_{1}}{g_{2}}} \cos \frac{z}{2} e^{- i \omega t} \\
\psi_{I_{2}}(z,t) &=& \sqrt{- \frac{g_{1}}{g_{2}}}\sin \frac{z}{2} e^{- i \omega t},
\end{eqnarray}
with the corresponding energy $E_{I} = \frac{\pi g_{1}}{6 g_{2}^{2}}\left[g_{1}^{2} - 3 g_{2} (2 + V_{0} - 2 \mu)\right]$.
In this case, the phase is arbitrary but the number of atoms on each lattice site is fixed. Thus, 
the atoms stop tunnelling between the lattice sites and the superfluidity is lost.
The filling fraction, in this case, is found to be $\nu_{I} = \frac{g_{1}}{2 g_{2}}[- 1 \pm \frac{\cos(n \pi) - 1}{n \pi}]$,
which is independent of the lattice potential and shows modulations similar to the superfluid phase.

\section{BEC with quintic nonlinearity} For tight transverse
confinement, strong interactions and low densities, the
  quintic nonlinear interaction term dominates the GP equation.  In
this case, we obtain only a superfluid phase with same periodicity of
the lattice potential: $\psi(z,t) = \sqrt{\sigma_{1}} e^{i \chi_{1} -
  i \omega_{1} t}$, where, $\sigma_{1} =
\left(-V_{0}/g_{2}\right)^{1/2} |\cos z|$ and $\omega_{1} =
\frac{1}{8} - \mu$.  Interestingly, this superfluid phase does not
have any background term. Amplitude of the lattice potential and the
three-body interaction strength need to be of opposite sign for the
solution to exist.  The supercurrent is found to be space-time
independent: $J_{1} = \sigma_{1} v_{1} = c_{1}$, where $c_{1} =
\frac{V_{0}}{8 g_{2}}$. Here, $v_{1}$ is the superfluid velocity:
$v_{1} = \chi'_{1} = \frac{c_{1}}{\sigma_{1}}$.  The filling fraction
is found to be a constant: $\nu_{1} = \frac{2}{\pi}
\left(-V_{0}/g_{2}\right)^{1/2}$.


\section{BEC with cubic nonlinearity} The GP equation, in
presence of pure two-body interactions ($g_{1} = g$) is known to
exhibit trigonometric solutions of the form \cite{Carr},
\begin{equation}\label{mono}
\psi(z,t)=\sqrt{\left(a-\frac{V_{0}}{g}\cos^{2}(z)\right)} e^{i
\chi_{2}(z)-i\omega_{2} t},
\end{equation}
where the background $a$ is a free parameter. The nontrivial phase,
$\chi_{2}(z) = \frac{c_{2} \tan^{-1}[\sqrt{1-\frac{V_{0}}{ag}}
\tan(z)]}{\sqrt{a(a-\frac{V_{0}}{g})}}$,
$\omega_{2} = \frac{1}{2}-\mu + ga$ and the integration constant $c_{2}^{2} =
\frac{(1 - 2 \mu - 2 \omega)(1 - 2\mu-2\omega + 2V_{0})}{4g^{2}}$.
The average atoms number density per lattice site is found to be related with
the background $a$: $\nu_{SF}=a-\frac{V_{0}}{2g}$. Here the
filling fraction is independent of number of lattice site $n$. The non-trivial phase
$(c_{2}\neq0)$ indicates the presence of superfluid component,
with a constant flow density $J_{2} = c_{2}$.
The potential, density and the phase have the same periodicity in this case, which is very different
from the previous cases, with two- and three-body interactions.
The ground state energy of the superfluid state
\begin{eqnarray}
E_{SF}&=&\frac{\pi
g}{8}\left(\frac{3V^{2}_{0}}{g^{2}}-\frac{8aV_{0}}{g}+8a^{2}\right)-
\frac{\pi c_{2}^{2}}{\sqrt{a\left(a-\frac{V_{0}}{g}\right)}}\nonumber
\\ &
& + \frac{\pi}{2}\left(2a-\frac{V_{0}}{g}\right) +
\frac{\pi}{2}\sqrt{a\left(a-\frac{V_{0}}{g}\right)}\nonumber
\\ & &-\pi\mu\left(2a-\frac{V_{0}}{g}\right)+\pi
V_{0}\left(a-\frac{3V_{0}}{4g}\right),
\end{eqnarray}
shows a branch cut at $a = a_{1} = 0$ and $a_{2} =
\frac{V_{0}}{g}$, where $a_{1T}$ and $a_{2T}$ are the critical values of
background. For $a_{1T}=0$,
$\omega _{2} + \mu = \frac{1}{2}$ and at $a_{2T} = \frac{V_{0}}{g}$,
$\omega_{2} + \mu = \frac{1}{2} + a g$; at these values the super-current ($J_{2}
= c_{2}$) vanishes and superfluid phase transits to the insulating
phase.
The energy is continuous; however, the first derivative at
the transition point (T),
is discontinuous like the previous case, hence, indicating that the phase
transition, is of first order. The
wave function in the insulating phase at $a_{1T}$ is, $\psi_{I_{1}}(z,t) =
\sqrt{\left(-\frac{V_{0}}{g}\right)}\cos z e^{-i(\frac{1}{2}-\mu)t}$, with the energy,
$E^{1}_{I}=\frac{\pi V_{0}}{8g}(8\mu-3V_{0}g-4)$. Corresponding number density is
$\nu^{1}_{I}=-\frac{V_{0}}{2g}$, indicating that in this insulating phase, the
interaction and potential strength have opposite signature. For the second insulating
phase at $a_{2T}$, $\psi_{I_{2}}(z,t) = \sqrt{\left(\frac{V_{0}}{g}\right)}\sin z
e^{-i(\frac{1}{2}-\mu + a g)t}$, with, $E^{2}_{I}=\frac{\pi V_{0}}{8 g}(5V_{0}g-8\mu+4)$,
and the average number density $\nu^{2}_{I}=\frac{V_{0}}{2g}$.

The interaction and potential strength, here, should have the same signature. For a positive $V_{0}$,
these two phases respectively belong to attractive and repulsive regimes. It is worth emphasizing that
for a fixed $V_{0}$ and coupling both these insulating phases only exist for a given number density of
atoms. In this case, as well as in the previous cases, interestingly we find that the dispersion affects
the density marginally, where it adds a constant term in the phase.

Keeping in mind the possibilities of charge density type of ground
states in one dimension, it is natural to investigate if other
type of solutions, still commensurate with lattice periodicity,
are allowed in this system. We have found such an exact solution
exhibiting \emph{rational} character:
\begin{equation}\label{solu}
\psi_{I}(z,t)=\frac{a+b\cos^{\alpha}(z)+c\cos^{\delta}(z)}{1+d\cos^{\beta}(z)}
e^{-i\omega_{3} t},
\end{equation}
with $\alpha=\beta=1$ and $\delta=2$. These solutions do not exist
in superfluid phase. The parameters for this insulating phase are
given by,
\begin{eqnarray*}
 a&=&-\frac{3}{4}\sqrt{-\frac{V_{0}}{g}}\mp\frac{1}{4}
\sqrt{\frac{12}{g}-\frac{9V_{0}}{g}}, b=\sqrt{-\frac{V_{0}}{g}},
\\
c&=&\pm\frac{V_{0}}{2}\left(\sqrt{\frac{12}{g}-\frac{9V_{0}}{g}}-3
\sqrt{-\frac{V_{0}}{g}}\right) \textrm{and} \\
d&=&\pm\frac{\sqrt{-V_{0}g}}{2}\left(-\sqrt{\frac{12}{g}-
\frac{9V_{0}}{g}}+3\sqrt{-\frac{V_{0}}{g}}\right),
\end{eqnarray*}
with, $\omega_{3} = \frac{1}{2}-\mu$. Here, the competition of the
lattice potential with the non-linearity is responsible for the
non-perturbative rational solutions with dual frequency character,
both commensurate with the lattice potential. It is worth
observing that, in case of solitons, non-linearity and dispersion
compensate each other, leading to stable localized solutions, as
well as periodic non-sinusoidal cnoidal waves. Presently, the
dispersion affects the character of the solution marginally. For
non-singular solutions, $V_{0}\leq-\frac{1}{6}$ in repulsive
interaction regime ($g>0$), and $V_{0}\geq\frac{4}{3}$ in
attractive regime ($g<0$). It should be noted that, in contrast to
the earlier found insulating phases, which are independent of the
background, the present density wave solution has a background,
which depends both on potential and the coupling. We further
observe that, both the modulations are necessarily present in the
presence of the lattice potential.

The energy function is non-analytic as $d \rightarrow0, \pm1$. For
$d=0$, the Pad\'e type of solutions representing insulating phase
reduces to the insulating phase exhibiting single frequency
behaviour found earlier. In the limit $d=\pm1$, the BEC is pushed
gradually to the unstable position on top of the hills of the
lattice potential indicating the unphysical nature of this
singular point.

\section{Experimental Realization} 
In the superfluid phase, the phase coherence between the different
sites of the optical lattice in a BEC has been established
experimentally through the formation of the interference pattern
between condensates located at the nodes of the laser standing
wave. In the insulating phase, the phase coherence is lost and hence,
no interference pattern is formed. The phases identified here, can be
detected in a method used by Cataliotti {\it et. al.}, in
\cite{Cataliotti} for pure two-body interaction. The density wave
insulating phase will not show phase coherence, however, unlike the
regular insulating phase, the spatial distribution of atoms will show
periodic behaviour commensurate with the lattice potential. The stripe
phase will show phase coherence between the atoms, where the atoms
will occupy only the alternating lattice sites.

To check the parameter regime for the possible existence of these type
of density wave and stripe phase, we consider the $^{87}Rb$ condensate
with $10^{4}$ number of atoms. Mass of the atom is $m = 1.44 \times
10^{-25}$ $kg$ and the transverse trapping frequency is
$\omega_{\perp} = 2 \pi \times 140$ $Hz$. The s-wave scattering length
$a = 5.4$ $nm$ \cite{Choi,Burt}. The two-body interaction strength is
given by, $g_{1} = 2 a m \omega_{\perp}/ \hbar k$.  The wave vector $k
= 2 \pi/\lambda = 8.06 \times 10^{6}$ $m^{-1}$. It is worth mentioning
that all the above parameters can be changed independently. Larger
values of interaction strengths can be achieved by higher density of
the condensate, higher $a$ and for small $k$. The scattering length
can be tuned as well by Feshbach resonance. Small $k$ may be achieved
by adjusting the relative angle between the two interfering laser
beams.  The optical lattice depth is taken as $V_{0} = 0.002$ (in the
units of $E_{r}$), implying that the potential is shallow, where the
mean field GP equation is valid. As we increase $V_{0}$, at some
point, phase coherence is lost and superfluidity breaks down and there
will be no tunnelling of atoms between consecutive lattice
sites. Hence, the phenomena discussed in this paper should be
observable within the present experimental capability.

\section{Conclusion} 
In summary, a superfluid density wave, having a periodicity twice that
of the lattice potential, exists in the presence of both
cubic and quintic nonlinear interactions, where the density
modulation is over a constant background. A stripe phase is present in
certain parameter regime, where superfluid matter is present in only
some domains of the lattice. These phases are absent, when, only cubic
and quintic nonlinear interactions are present. This density wave
phase transits to an insulating phase through DSIT.  The phase
coherence is lost due to the modulational instability. The stability
analysis of this solutions under VK criterion, supports the
development of modulational instability in this system.  The filling
fraction is found to vary inversely with the lattice location from a
given site, and differs for odd and even $n$. This different behaviour
of the filling fraction in the neighbouring sites can be traced to the
presence of the both cubic and quintic nonlinearities and
is absent when pure cubic or quintic nonlinear interactions
are operative.  In pure two-body case, the periodicity of the density
is same as the lattice potential.  A density wave insulating phase is
observed in this case, having two frequency modulations, commensurate
with the lattice potential. Explicit calculations show that these type
of density wave states and phase transitions can be probed in the
present laboratory conditions.

\section{Acknowledgement}
The authors acknowledges Ashutosh Rai for many useful discussions.

\end{document}